\documentstyle[iopconf1,epsfig]{article}
\begin{document}
\title{Status and Perspectives of Indirect Search for Dark Matter}
\author{Olga Suvorova}

\affil{\ Institute for Nuclear Research, 
Russian Academy of Sciences,  
60th October Anniversary Av. 7a,
Moscow, 117312, Russia}

\beginabstract
I summarize here and discuss the results of
presently operating neutrino 
telescopes in searching for a singnal of dark matter weakly interacting 
massive particles (WIMPs).
\endabstract

\section{Introduction}
The status of dark matter problem is quite complicated (for a review see 
ref.\cite{ref:prim,ref:jung}) but clear displays the close connection of
the 
fundamental characters of the early Universe with the interaction properties 
of elementary particles. Existing evidence of a cosmologically significant 
amount of dark matter being able to explaine as much as 99\% of the mass of 
the Universe together with Big Bang Nucleosynthesis and formation of large 
structures in the Universe supports hypothesis of the non-baryonic and 
mainly cold (75\%) dark matter bulk.

In a Minimal Supersymmetric Standard Model (MSSM) with R-parity conservation 
the lightest of the four neutralinos assumed to be stable against decay and
annihilating at weak scale \cite{ref:kane}. Thanks to that neutralino provides 
a relic density close to the critical one.
 
The opportunity for indirect signature as well as for direct detection of WIMPs
is coming from a drastical assumtion that WIMPs elastically interact (via 
mass and spin) with ordinary baryonic matter \cite{ref:kane,ref:frees}. Since 
neutralinos are a mixture of the respective partners of the electrically neutral 
gauge bosons B and $W_3$ (gaugino states) and two Higgs bosons (higgsino states), 
a value of the neutralino mass under demand of a grand unification 
\cite{ref:nath1,ref:nath2}
is defined by four independent space parameters: $m_0, m_{1/2}, A_t, tan{\beta}$ 
and also $sgn(\mu)$ in its convensional means. The experimental lower mass 
limits have been set by large accelerators \cite{ref:acce} ($ > 25.9 eV$), 
while an upper mass limit is assumed to be in TeV range respectively to a 
suggestion about neutralino cosmological aboundace. 

The underground neutrino telescopes have measured in global an order of three 
thousands of the muonic neutrino events from down hemisphere. Statistics allows
to look for an excess of upward-going muons relatively to the expectations 
from atmospheric neutrinos in the directions of the Sun and near the centre of 
the Earth, while these astrophysical bodies could be considerated as energetic
neutrino sources. Accordinally to the theory in the solar system the capture 
rates of neutralinos by the Sun and the Earth are enough large to provide a 
thermal equilibrium with the annihilation rate at the scale of the Universe
age \cite{ref:seck} inside its central regions where neutralinos might get 
accumulated due to crossing gravitationally trapped orbits 
\cite{ref:frees,ref:gold} and subsequent energy losses via elastical scattering
\cite{ref:ress}. 
Decays of generated fermions and bosons in neutralino annihilation channels 
produce high energy leptons from which only neutrinos survive on the way to 
the detector. 

Presently neutrino experiments of the Baksan \cite{ref:bakDM96} and MACRO 
\cite{ref:MAC99} following IMB \cite{ref:IMB} and Kamiokande \cite{ref:KAM}
researches of dark matter WIMPs have set a more stronger muon flux limits at 
the 90\% confidential level (c.l.), since they did not measure significant 
excess of upgoing muon events in the direction of the pointed out predicted 
neutralino sources.

As it has been shown \cite{ref:bott}, a detector with an area of 
$km^2$ running for one year would be sensitive to a few TeV range of neutralino 
masses. The first results from Baikal \cite{ref:BAI1,ref:BAI2} and AMANDA 
\cite{ref:AMA1,ref:AMA2} as well as demonstrator of ANTARES project 
\cite{ref:ANT1} presenting principly new generation of neutrino detectors 
offer having more sensitive instruments for indirect searching for WIPMs.

\section{General considerations of the method}
The efforts in searching for neutralino indirect signature aim to differentiate 
the measured flux of neutrino events into expected one from neutralino sources 
and from atmospheric neutrinos background. The expected number of upward-going 
muons is defined by a fall of neutrino flux ($dN_{\nu_j}/dE_{\nu}$)
from a source at distance ($R$) and by a probabilty ($P_j(E_{\nu})$) that j-th type 
neutrino with primary energy $E_{\nu}$ will produce a muon which arrives at the 
detector with energy $E_{\nu}$ higher than threshold $E_{th}$. For a remote 
moving source of neutrinos (as the Sun) the detector rate depends also on a 
track on the sky. The value $P(E_{\nu})$ is determined by second momentum of 
the $E_{\nu}$, because of the $\nu_{\mu}N$ cross section and muon path range 
increase with neutrino energy. The spectrum of atmospheric  $\nu_{\mu}$-background 
is a decreasing $E^{-3}$ function of $\nu$-energy. Hence, for a more energetic 
muons it is more advantageously to look for a DM signal from predicted neutrino 
sources, including the Earth core and the Sun. Due to the neutralino mass value
which allowes to have one order magnitude difference, the generated neutrino 
fluxes from decay processes of neutralino annihilation products will be 
characterized by different hardness of spectrum respectively to a branching 
ratios ($B_i$) into annihilation channels:
$$
{{dN_{\nu_j}} \over {dE_{\nu}}}={{\Gamma_A} \over {4 \pi R^2}}
\sum_i B_i{{dN^i_{\nu_j}} \over {dE_{\nu}}}, \qquad 
\nu_j=\nu_{\mu},{\bar \nu}_{\mu}.
$$
Here $dN^i_{\nu_j}/dE_{\nu}$ is the differential spectrum of j-th neutrinos at the
surface of the the Sun or the Earth produced in annihilations 
into the i-th channels. While the branching ratios are calculated accordinaly
to the theory of annihilation cross sections \cite{ref:drees}, the annihilation
rate ($\Gamma_A$) is evaluted from capture rate and is a function depending on a
few astrophysical and nuclear values which are known with large uncertanties 
\cite{ref:jung,ref:nath2,ref:ress}. 
Notice that the Baksan collaboration has obtained \cite{ref:bakDM96} 
the 90\% c.l. upper limits on annihilation rate defined to be completely detector
independent and as a function of the neutralino mass. They could be
directly 
applied to comparison with a theoretical calculations, in contrary to the 
limits on muon fluxes being detector depending values. 

The modern telescopes do not measure neutrino spectra. So discrimination of 
nonatmospheric origin neutrinos from atmospheric ones is carried out on
zenith 
angular distributions. The angular resolution is dominated for this task 
together with the existing detector's acceptance for different neutrino
fluxes.

In the case of neutralino sources the expected signal is strongly collimated.  
Fig.1 shows the response of Baksan underground scintillator telescope on Monte 
Carlo simulated neutralino annihilations inside the Earth core and the Sun 
for different masses of neutralinos. 

Although the distributions reflect the installation construction and depend on 
neutralino composition, nevertheless it seen that a more larger neutralino 
mass gives a more narrow shape of the distribution. Optimization of the 
signal to background ratio by appropriate angular selection of data on basis 
of signature of modeled neutralino annihilations has been done by the Kamiokande,
Baksan, MACRO and AMANDA collaborations. Tables 1 and 2 summarize the results 
over the operating telescopes, displaying the 90\% flux limits and search 
half-cones collecting 90\% of expected neutralino signal. The angular maximum 
is followed to be not higher than $25^\circ$ for the Earth and $16^\circ$ for 
the Sun. Under the LEP bound \cite{ref:acce} on neutralino mass these angular 
window limits are decreasing at least on $5^\circ$.

\section{Simulation models}
There are a few theoretical groups \cite{ref:jung,ref:gond,ref:bott} 
which carried out SUSY models down to an applicable neutrino spectra from 
neutralino annihilation for further calculation of upgoing muon fluxes at 
a given neutrino telescope. All of them contain a different incorporated 
effects related with the calculation of relic density of neutralinos, so 
below the authors of SUSY models will be indicated.     
\begin{figure}
\centering
\mbox{\epsfig{file=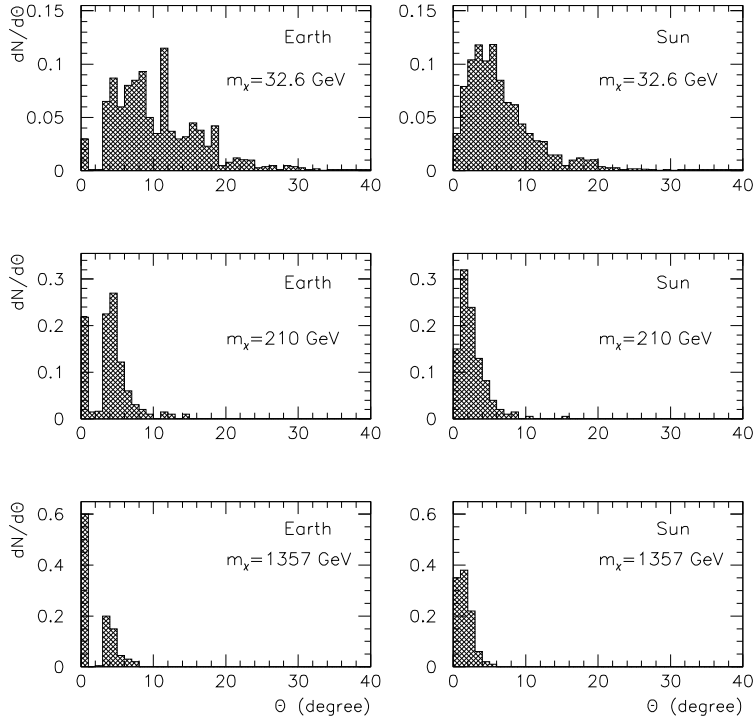,height=11.0cm}}
\caption
{Zenith-angular distributions of MC simulated upward-going muons 
from $\chi-\chi$ annihilations. Parameter set: $tan\beta=8$, gaugino
part is 0.5, pseudo-scalar Higgs mass $m_A$=121 GeV,$ \mu>0.$}
\end{figure}
Simulation of neutrino signal from neutralino annihilation implies calculation
of neutralino space spread in the source, which is determinated by local 
gravitational potential. With an approach of constant neutralino temperature 
equal to the central one, neutralinos have spaced with a shape of Gaussian 
distribution with inverce dependence on mass \cite{ref:seck,ref:bott}. While 
in case of the Earth predicted angular distribution of annihilation neutrinos 
is strongly correlated with the neutralino space distribution, inside the Sun 
the distribution of generation points of neutrino have to be accounted in the 
sense that there are energy losses both in quarks hadronization processes and 
in neutrino passage through a high density layes of the Sun resulting to 
neutrino spectrum modification \cite{ref:RS}. The neutrino's yield 
($dN^i_{\nu_j}/dE_{\nu}$) is evaluting on basis of inclusive neutrino spectra
arising in simulations of $e^+e^-$ annihilations according 
to PYTHIA versions and with applying of two-body decay kinematics to boson and
t-quark channels of neutralino annihilations (see for review \cite{ref:jung}).
\begin{table}
\begin{center}
\footnotesize\rm
\caption{The 90\% c.l. upper limits on upward-going 
muon fluxes for ${\chi}{\bar {\chi}}$ annihilations in the Earth's
core and half-cones collecting 90\% of $F^{ann}_{\mu}$}
\begin{tabular}{p{7mm}p{6mm}p{9mm}p{6mm}p{10mm}
p{10mm}p{6mm}p{10mm}p{10mm}}
\topline
{ }&\multicolumn{2}{c}{BAKSAN} 
&\multicolumn{2}{c}{MACRO} 
& \multicolumn{1}{c}{BAIKAL} 
& \multicolumn{3}{c}{AMANDA B10} \\
\bs
{Half-cone} & ${m_{\chi}}$ 
& {$\Phi^{upper}_{\mu}$} 
& {$m_{\chi}$} 
& {$\Phi^{upper}_{\mu}$} 
& {$\Phi^{upper}_{\mu}$} 
& {$m_{\chi}$} 
& {$\Phi^{hard}_{\mu}$} 
& {$\Phi^{soft}_{\mu}$} \\
{(${}^\circ$)} & {(GeV)}
& {($10^{-14}$ cm${}^{-2}$s${}^{-1})$}
& {(GeV)}
& {($10^{-14}$ cm${}^{-2}$s${}^{-1})$}
& {($10^{-14}$ cm${}^{-2}$s${}^{-1})$}
& {(GeV)}
& {($10^{-14}$ cm${}^{-2}$s${}^{-1})$}
& {($10^{-14}$ cm${}^{-2}$s${}^{-1})$} \\
\midline
30.0 & & & & 2.01 & 11.0 & & & \\
25.0 & 12.8 & 3.2  & & & 9.3 & & & \\
24.0 & & & & 1.56 & & & & \\
18.0 & 32.6 & 2.1 & & 1.28 & 5.9 - 7.7 & & & \\
15.0 & & & & 1.03 & 4.8 & 100 & 14.1 & 103.8 \\
 & & & & & & .... & & \\ 
 & & & & & & 5000 & 2.1 & 3.4 \\
14.8 & & & 60 & & & & & \\
11.8 & & & 100 & & & & & \\
10.0 & 82.8 & 0.93 & & & & 100 &  9.8 &  82.2 \\
 & & & & & & .... & & \\ 
 & & & & & & 5000 & 1.3 & 2.1 \\
9.0 & & & & 0.658 & & & & \\
8.4 & & & 200 & & & & & \\
7.0 & 210 & 0.62 & & & & & & \\
6.0 & & & 500 & 0.507 & & & & \\
5.0 & 535  & 0.54 & & & & 100 & 23.6 & 231.3 \\
4.9 & & & 1000 & & & 250 & 3.7 &  16.4 \\
4.5 & 1358 & 0.52 & & & & 500 & 2.3 & 6.2 \\
4.0 & 3454 & 0.52 & & & & 1000 & 1.8 & 3.8 \\
3.0 & & & & 0.289 & & 5000 & 1.3 &  2.3 \\
\bottomline
\end{tabular}
\end{center}
\end{table}
Monte Carlo simulation of neutrino signal at the detector subdivides in two 
parts. At first one, neutrino interactions with the matter surrounding the 
detector and muon propagation up to the detector are modeling. Neutrino flux
being calculated in dependence on model (atmospheric origin or neutralino 
source) is inserted into Monte Carlo generator. At the day, the Baksan and 
MACRO calculations have the same simulation scheme 
\cite{ref:baktaup97,ref:MAC2} 
which applied the Bartol atmospheric $\nu$ flux \cite{ref:ansb95}.

The second part means calculation of acceptance for the arrived muons with the
same requirements for hardware triggers and the same set of a cuts as for real
data for exclusion of those of neutrions candidates which could be mimicked by
particles produced by downward-going atmospheric muons with the large scattering
angles or multiple muons (see \cite{ref:baktaup97,ref:MAC2}). 

There are a running routines of reproduce Cherenkov technigue registration
of relativistic neutrino-induced muons passing through a volum of pure water 
(the Super-Kamiokande) or natural water (the BAIKAL, the ANTARES project) or 
ice (the AMANDA) medium.

The detail descripsion of installations are outside the scope of this paper,
but for DM searching it is important to note a few characters of operating
neutrino detectors and the obtained ratios of measured events to expected one.

\begin{table}
\begin{center}
\footnotesize\rm
\caption{The 90\% c.l. upper limits on upward-going 
muon fluxes for ${\chi}{\bar {\chi}}$ annihilations in the Sun
and half-cones collecting 90\% of $F^{ann}_{\mu}$}
\begin{tabular}{p{10mm}p{14mm}p{16mm}p{14mm}p{16mm}}
\topline
{ }&\multicolumn{2}{c}{ BAKSAN} &\multicolumn{2}{c}{ MACRO} \\ 
\bs
{Half-cone} & ${m_{\chi}}$ 
& {$\Phi^{upper}_{\mu}$}
& {$m_{\chi}$} 
& {$\Phi^{upper}_{\mu}$} \\
{(${}^\circ$)} & {(GeV)}
& {($10^{-14}$ cm${}^{-2}$s${}^{-1})$}
& {(GeV)} 
& {($10^{-14}$ cm${}^{-2}$s${}^{-1})$} \\
\midline
30.0 & & & & 6.38 \\
24.0 & & & & 4.06 \\
18.0 & & & & 2.77 \\
16.0 & 12.8 & 2.4 & & \\
15.0 & & & & 2.07 \\
13.0 & 32.6 & 2.1 & & \\
12.9 & & & 60 & \\
10.3 & & & 100 & \\
9.0 & & & & 1.42 \\
7.5 & & & 200 & \\
6.5 & 82.8 & 1.1 & & \\
6.2 & & & 500 & \\
6.0 & & & & 1.07 \\
5.8 & & & 1000 & \\
4.7 & 210 & 0.65 & & \\
3.6 & 535 & 0.64 & & \\
3.0 & 1358 & 0.64 & & 1.35 \\
2.7 & 3454 & 0.64 & & \\
\bottomline
\end{tabular}
\end{center}
\end{table}

\section{Detectors: operating and running in progress}
\subsection{Scintillator detectors: BAKSAN and MACRO}
Neutrino experiments at the Baksan telescope \cite{ref:bakICRC79} under 
effective rock thickness of 850$hg/cm^2$ and at the MACRO \cite{ref:MAC2} 
(3700 m.w.e. depth), are realized on basis of large scintillator detector and 
discrimination upward and downward-going muons by means of the time-of-flight
method \cite{ref:bakNT91}. The MACRO detector have a few better parameters for a 
detection of muons from down hemisphere, first of all thanks to its strong 
reduction of background from the downward atmospheric muon flux, by factor of 
$10^6$. While the Baksan location is relevant to the $5\cdot10^3$ reduction. 

Trajectories of a penetrating particle at the Baksan are determined by the 
positions of hitted tanks, which put together a system of 3,150 liquid 
scintillator counters of standard type $(70cm \times 70cm \times 30cm)$
which entirely cover eight planes of fourfloor telescope's building 
$(17m \times 17m \times 11m)$. The configuration provides $2^{\circ}$ of angular 
accuracy and the mean muon energy of about 20 GeV. The MACRO apparatus includes
the system of around 20,000 $m^2$ of streamer tubes, that able to get a 
track's reconstruction within $.5^\circ$, although the angular accuracy gets 
worse for inclined tracks mainly contributed in the analysis of muons pointing
in the direction of the Sun, due to installation's configuration 
($12 \times 76.6 \times 9 m^3$).

The Baksan and the MACRO detectors are sensitive to searching for neutralinos 
with the masses from the lower limit \cite{ref:acce}. That is because of theirs
low energy threshold: $E^{\mu}_{th} \approx 1 GeV$ for vertical upward 
through-going muons. Presently, the both expriments have reached the same order 
of statistics, hence the results could be comparable. 

During the Baksan observation from December 1978 until November 1998 the rejection 
of downward-going particles by hardware triggers left of about $6,5\cdot10^6$ 
events, while the triggers provides $0.1\%$ of the initial rate. Subsequent 
requirement of negative value of $1/\beta$ around -1 (that relevants to 
upward-going particle) has selected 1056 events. There are additional cuts
to 
a single track in order to exclude events mimicked by downward-going particles
\cite{ref:bakDM96,ref:bakNT99}. The Baksan gave an evidence of 713 upward 
through-going muons with 14.8 years of live-time \cite{ref:bakNT99}. The last 
MACRO update \cite{ref:MAC99} of $\nu$-events observation from March 1989 
until February 1999 respects to 3.93 live-years in MACRO full configuration. 
Two different topologies of the MACRO detector \cite{ref:MAC2} are used in the
$\nu$-registration: upward through-going muons and internally produced 
upward-going particles, the energy range of latter one is close to $E^{\mu}_{th}$.
Therefore, a sample of 642 upward through-going muons have been applied for 
searching for neutralinos from the Earth's core and 971 upward-going muons 
(including semi-contained events) are in case of the Sun.

Expectations at the Baksan from Monte Carlo simulations of atmospheric neutrinos 
(20 runs of live-time) in total number of events agree with measurements and 
in particular, there is consent in vertical bins within error bars, as it 
seen from Fig.2. The ratio of measured to expected events is found to be 
$0.95\pm0.04(stat.)\pm0.08(syst.)\pm0.15(theor.)$. The MACRO experiment has 
found a significant deficit of events in vertical bins in comparison with 
expectations from atmospheric neutrinos \cite{ref:MAC2}. The pointed out ratio
have been obtained to be 
$0.740\pm0.031(stat.)\pm0.044(syst.)\pm0.12(theor.)$ \cite{ref:MAC99}. 

However, the fits of the shape of the experimental zenith angular distribution 
by the predicted one from the theory both with neutrino flavor oscillations 
and without one have a poor probabilities at the Baksan as well at the MACRO 
detectors \cite{ref:baktaup97,ref:mikh,ref:MAC2,ref:MAC99}. 

In the direction of the Sun both the Baksan and the MACRO experiments have 
concluded the absence of additional neutrino source relatively to the background 
evaluated from data itself \cite{ref:baktaup95,ref:bakDM96,ref:MAC99}.
The Baksan distribution of angles between the muon arrival direction and a 
radius-vector of the Sun is presented in Fig.2. There is a comparison with the 
measured background derived from data correlation with positions of fake 'suns' 
(points on the sky with the same acceptance as true Sun but shifted at some 
hour angles). 
 
With the same order of statistics and similar defined energy thresholds
the MACRO and the Baksan experiments have set a very close limits on muon fluxes 
at the 90\% c.l. (see Tables 1 and 2). The angle ranges collecting 90\% of 
neutralinos signal as a functions of its value of mass have been obtained 
with different SUSY models. Notice that the flux upper limits presented by the 
MACRO experiment does not connect here with found half-cones. The Baksan 
results have been derived \cite{ref:bakDM96} 
on the basis of phenomenological approach to MSSM and Monte Carlo simulations 
in studying of two detector dependent values: a muon detection probability per 
one neutralino pair annihilation both for the Earth and the Sun cases and
a half-cones containing 90\% of events. Therefore, the energy dependence of 
the detector acceptance on fall neutrino flux from different neutralinos 
has been accounted. Monte Carlo simulations have reproduced of 1,000 arrival 
muons from each set of neutralinos. The optimization of searching half-cones for 
neutralino signal at the MACRO detector has been done with SUSY models of 
Bottino et al.\cite{ref:bott}. 
\begin{figure}[htb]
{\psfig{figure=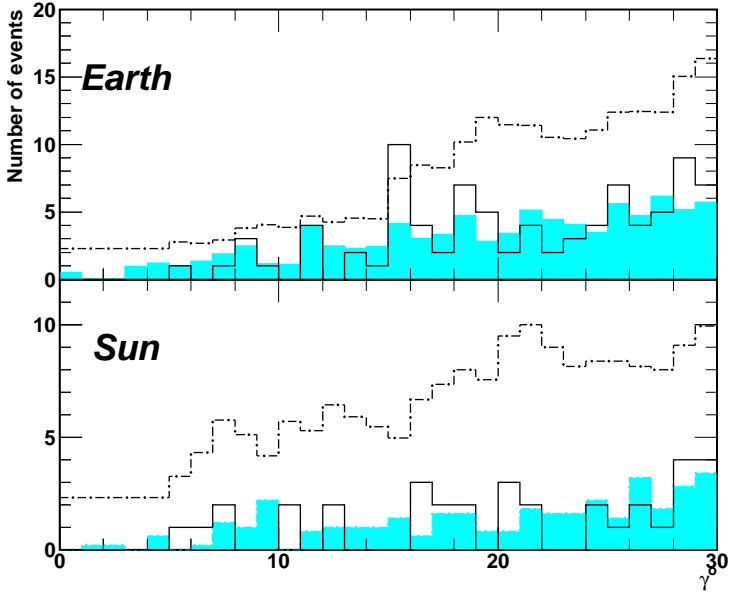,height=8.0cm}}
\protect\caption[]
{Distribution in $\gamma$ for upward-going muons from directions of the
Earth's core and the Sun. Solid histogram is data and shaded one is
expectation for atmospheric neutrinos. Dash-dotted histogram is upper limits 
at 90\% c.l. on number of upward-going muons produced by neutrinos 
nonatmospheric origin as a function of cone opening angle $\gamma$.}
\end{figure}

In determination of conservative limits on muon fluxes from the direction
of the
Earth's core, the MACRO has assumed equality between the number of measured 
events and the number of expected one after applied normalization separately 
to each of the considered search cones \cite{ref:MAC99}: from $3^\circ$ to 
$30^\circ$. The error of 5\% has been estimated as maximum \cite{ref:MAC99} 
with respect to accounting the dependence on energy of acceptance of the MACRO 
detector with changing of neutrino energy spectra from different neutralino 
masses. 

The obtained 90\% c.l. flux limits as a function of neutralino mass have ruled 
out a considerable number of SUSY parameter space for the Sun as well as for 
the Earth, including one related with the interpretation of the DAMA/NaI data 
\cite{ref:DAM}. Notice that the MACRO determination of exclusion regions 
\cite{ref:MAC99} assumed a rescaling of the mass density of the neutralino 
\cite{ref:bott}, which directly affects the capture rate, subsequently, muon 
flux value and hence, the conclusion on ruled out space.
 
The Baksan detector ability to probe MSSM parameter space has been shown
previously in Ref.\cite{ref:baktaup95}, where a large part of the space 
covering the resonance mass range for the Earth's capture \cite{ref:gold}
has been ruled at 90\% c.l. as well as many SUSY models calculated by 
Gondolo et al.\cite{ref:gond}. The recent analysis of Bergstrom et al.
\cite{ref:berg} with contribution from a new population of WIMPs coming from 
Galactic halo WIMPs \cite{ref:kraus} which could be scattered in the outer 
layers of the Sun, has demonstrated a sizeable enhance the neutralino signal 
from the Earth. The conclusion is that many more models around 60-130 GeV 
previously thought to be allowed by Baksan but above the claimed DAMA 
bound \cite{ref:DAM} should really be considered as ruled out with even 
greater confidence. 

\subsection{Cherenkov detectors: BAIKAL, AMANDA and ANTARES}
The method of registration of Cherenkov light emitted by passing relativistic 
neutrino-induced muon in the large natural volume of water or ice is performed 
by a three-dimensional matrix of optical modules (OMs), which is deploying at 
the depth.

The pioneer realization of the BAIKAL running project in deployment of 72 m 
long strings supporting OMs at 1,100 m depth \cite{ref:BAI2} and successful 
deployment of kilometer-long strings with OMs to 2300m depth at the South Pole
by AMANDA collaboration \cite{ref:AMA2} have provided observation of muons 
induced by very energetic muonic neutrino and thus, started the new 
generation of $km^3$ detectors.  

Both experiments have began in 1993 and have carried out a few season installations
on basis of the complex programms with monitoring of the properties of environment 
used for a particle detector and a lot of tests and calibrations in studying of
transparency, radioactivity, the scattering length. The reduction of the 
biofouling deposited on the glass spheres with a time is the main care
both 
at the Baikal and at a deep-sea running project of the ANTARES collaborations 
\cite{ref:ANT1} started in the spring of 1996. Currently the ANTARES collaboration 
aims at building a $0.1 km^2$-size detector made of 500 m long strings at 2350 m 
depth of Mediterranean sea by the end of 2000. 

The operating large Cherenkov detectors have demonstrated that there is no 
up/down confusion of nearly vertical upgoing muons. Presently, the Baikal
experiment has found 4 candidates survived all trigger cuts (energy
threshold 
is order 10 GeV) with 70 live-days of NT-96 detector \cite{ref:BAI1} which
overlapping around $1,000 m^2$. The 90\% c.l. flux limits 
in the direction of the Earth's core \cite{ref:BAI1} are present in Table 1. 
The AMANDA collaboration gave an evidence of 11 upward-going muons in angular 
window $165^\circ$ with 85 days of AMANDA-B10 detector live-time from analysis 
of half of 1997 data sample \cite{ref:AMA1}. They estimate that it can pinpoint 
the direction of a high-energy muon with an uncertainty of about $3^\circ$.
The AMANDA preliminary statement is that \cite{ref:AMA1,ref:AMA2} both muon flux 
and angular distribution of upgoing events agree with expectations from Monte 
Carlo simulations of atmospheric neutrinos (three years of live-time) within 
error bars. For near 
vertical contained events an energy threshold might be estimated not much 
than 25 GeV. Upper limits on muon fluxes coming from the centre of the Earth 
have been set at 90\% c.l. for the neutralino mass range 100 GeV - 5 TeV with 
separation into soft and hard annihilation channels \cite{ref:AMA1} (see 
summarized results in Table 1). This idea is a very fruitful, since the detector
sensitivity is essentially improving due to adequate determination of the 
detector acceptance. 

The measurement of the arrival time of the Cherenkov light by OMs means both a
reconstruction of the muon direction and estimation of its energy. Due to 
optimization of the planning geometry of the ANTARES detector performed by 
Monte Carlo simulations \cite{ref:ANT1} it has been inferred two energy 
different possibilities for the detection. First is measurement of fully 
contained muon tracks at the detector. Respective muon energy range would be 
5-60 GeV that is suited for the neutrino oscillation study and the neutralino 
search. At higher muon energy ($>TeV$) a precision of reconstruction of the 
energy spectrum gets worse into 2-3 times because of fluctuation of detected 
number of photons. In this case found geometry is relevant for study of the 
very high energy neutrino astronomical sources ($>100 TeV$ at $0.1km^2$) with 
a very good angular resolution: better $0.2^\circ$ \cite{ref:ANT1}.

The case of ideal detector operating as a separate string of 500 m height 
with
about 100 OMs grouped by floors to use only full contained events coming from
nadir has been analyzed by F.Blondeau \cite{ref:ANT2}. The expectation of 22
upgoing muons from atmospheric neutrinos has been obtained per one 
$string \cdot year$ exposition. Detector response to neutralino signal from 
the Earth core has been tested with Neutdriver code \cite{ref:jung}. Obtained
detector sensitivity around 60 GeV neutralino mass is seen in Fig.3 where 
the regions of MSSM parameter space allowed by the ANTARES detector
\cite{ref:ANT2}
are shown in comparison with excluded counters from the Baksan \cite{ref:baktaup95}.
\begin{figure}
\centering
\mbox{\epsfig{file=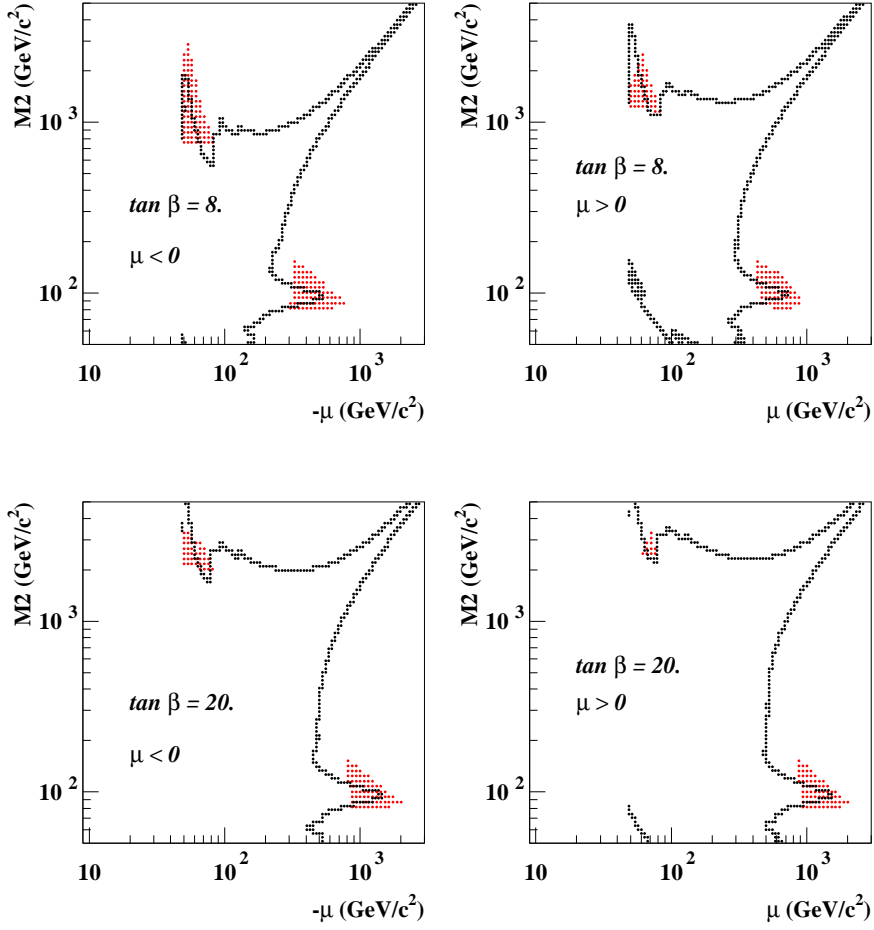,height=14.0cm}}
\caption
{The ANTARES allowed regions together with excluded one by the Baksan limits on the upward-going muons at the 90\protect\% c.l.} 
\end{figure}

\section{Perspectives}
The common perspective for the indirect searching for WIMPs in high energy
neutrino experiments is connected with a quantity and a quality of measured 
statistics of muons coming from down hemisphere. There is also a dependence on 
technique of Monte Carlo simulations in complete reconstruction of neutrino 
signal accordingly to its origin. Finally, present uncertainties of 
astrophysical and nuclear parameters which determinate neutralino capture rates 
as well a priori unknown preferable region of space parameters could not give 
an unambiguous bound on mass and composition of neutralino even by planning
detectors of upward-going muons.  

Among the operating underground telescopes Super-Kamiokande \cite{ref:SK} 
gives a very impressive results with high statistics confirming the 
zenith-angular-dependent deficit for contained events, to which atmospheric 
neutrinos in the 0.1-10 GeV energy range are relevant. Having in mind the 
last update of upward through-going muons with 1021 data sample per 923 days 
\cite{ref:SK} corresponding to $E_{th}$ order 7 GeV, the expected signal 
from the neutralino with masses covered resonance range for the Earth could be 
tested. The recognition of neutrino flavor oscillations would play a crucial 
role, since this process could decrease the expected neutralino signal 
considerably (by factor 0.5-0.8 for neutralino mass less 100 GeV) as it has 
been shown by Fornengo \cite{ref:forn}. 
Notice that there is a $\tau$-lepton channel in the neutralino annihilations 
which offers a generation of $\nu_{\tau}$ in decay chain and a hard spectrum 
of $\nu_{\mu}$.

In perspective one may wait both the resolving of the measured deficit of 
nearly vertical upward-going muons at the MACRO detector and determination 
more precisely the influence of instability of the detector parameters during 
long term operation of the Baksan telescope, since the obtained systematic 
uncertainties are too large (8\%). The data of the Baksan and the MACRO with 
relatively low energy threshold (1 GeV) could be used to set a separate limits 
on soft and hard neutrino fluxes and as to be compared with accelerator limits 
in SUSY parameter space.

The running projects of new generation of neutrino telescopes imply a 
kilometer sizes, but with larger effective area they will have also higher 
energy threshold and thus, could lose a sensitivity to the neutralino mass 
range (50-100 GeV) where is expected a resonance WIMP capture in the Earth.
The detail discussion of trade-off between area and energy threshold for 
indirect search for dark matter by km-size detector are presented in 
Ref.\cite{ref:gond}. The method using full contained events coming near nadir 
at the ANTARES project developed for neutrino oscillation study is predicted 
to detect a signal from neutralinos with a mass close to 60 GeV.

\section{In conclusion}
The operating neutrino telescopes did not gave an evidence for additional
sources of high energy neutrinos from the centre of the Earth as well from 
the Sun, and the 90\% c.l. limits on upward-going muons fluxes are setten.

The indirect method of searching for WIMP signal directly depends on value of
uncertainties both measurements and theory, the decreasing of which is too 
complicate task. The optimization of the ratio of expected neutralino signal 
to background from atmospheric neutrinos is the applied way in searching for
dark matter WIMPs. There is a complementarity of the low and high energy 
threshold detectors which are sensitive to a different neutralino mass range.

\section*{Acknowledgments}
I am grateful to the organizers of the Beyond99 and Prof. H.V.Klapdor-Kleingrothaus
for inviting me at the Ringberg Castle and for the warm hospitality. Special 
thanks are given to colleagues at the BAKSAN, Drs. M.M.Boliev, A.V.Butkevich 
and Prof. S.P.Mikheyev for collaboration and efforts made for a long time.

\end{document}